\documentclass[aps,nofootinbib,amsfonts,groupedaddress]{revtex4}
\usepackage{amsmath}
\usepackage{epsfig,epsf}
\usepackage{graphicx}
\usepackage{verbatim}
\usepackage{epstopdf}
\usepackage{bm}  
\usepackage{epsfig,epsf}
\def\beq{\begin{equation}}
\def\eeq{\end{equation}}
\def\bea{\begin{eqnarray}}
\def\eea{\end{eqnarray}}
\def\eq#1{{Eq.~(\ref{#1})}}
\def\fig#1{{Fig.~\ref{#1}}}

\newcommand{\Lb}{\left(}
\newcommand{\Rb}{\right)}
\parskip=\itemsep               

\newcommand{\pom}{I\!\!P}

\def\pom{{I\!\!P}}

\begin{document}

\begin{flushright}
TAUP 2974/13
\end{flushright}

\title{Proton-air collisions in a model of soft interactions at high 
energies }

\author{E.~ Gotsman$^{1}$,~ E.~ Levin$^{1,2}$ ~and~ U.~Maor${}^{1}$}

\affiliation{
${}^{1}$~Department of Particle Physics, School of Physics and Astronomy,
Raymond and Beverly Sackler
Faculty of Exact Science, Tel Aviv University, Tel Aviv, 69978, Israel\\
${}^2$~Departamento de F\'\i sica, Universidad T\'ecnica Federico Santa 
Mar\'\i a, Avda. Espa\~na 1680\\ and Centro 
Cientifico-Tecnol$\acute{o}$gico de Valpara\'\i so,Casilla 110-V, 
Valpara\'\i so, Chile}%

\date{\today}

\begin{abstract}
  We show that Pomeron interactions generate important corrections to the
Gribov-Glauber formula, which is used to extract proton-proton cross
sections from proton-air  collisions at high energy. We show that these
corrections are larger than the errors for proton-air cross sections
measured at ultra high energies in cosmic ray experiments. We present a
description of these data in our model for soft interactions at high
energies, which describes all available accelerator data including that
from the LHC.
\end{abstract}

\maketitle
\section{Introduction}
\par
In this letter we examine the problem of hadron-nucleus interaction at high 
energies. It is
well known that in the Gribov-Glauber approach \cite{GLAUB,GRIBA}, 
where
the total cross section of the hadron-nucleus interaction is expressed 
through the inelastic cross section of hadron-proton scattering, can only 
be
 justified  at rather low energies, where the corrections due to Pomeron
 interactions may  be neglected. A more general approach has been
 developed\cite{SCHW,KAID,BGLM,GLMAA} in which the Pomeron interaction
 has been taken into account in the high energy range for :
\bea \label{KRHA}
&&g_{i}\,S_{A}(b)\,G_{3\pom}\,e^{\Delta_{\pom} Y} \,\,\propto\,\,g\,
G_{3\pom} A^{1/3}\,e^{\Delta_{\pom} Y} \,\,\approx \,\,1;\nonumber\\
&&\,\,\,\,\,\,\,\,\,\,\,\,\,\,\,\,\,\,\,\,\,\,\,\,\,\,\,\,\,\,\,\,
G^2_{3\pom}\,e^{\Delta_{\pom} Y} \,\,\ll\,\,1.
\eea
  $G_{3\pom}$ is the
triple
Pomeron coupling, $g$ is the vertex of Pomeron nucleon interaction, and
 1 + $\Delta_{\pom}$ denotes the Pomeron intercept.
 For the   nuclear profile $S_A(b)$ we use the general 
expression
\beq \label{SA}
S_A(b)\,\,=\,\,\int^{+ \infty}_{- \infty}\,d z\,\rho\Lb z,b\Rb\,\,=\,\,\int^{+ \infty}_{- \infty}\,d z\ \frac{\rho_0}{1\,\, +
\,\,e^{\frac{\sqrt{z^2 + b^2} - R_A}{h}}}; \,\,\,\,\,\,\,\int d^2 b\,S_A\Lb b \Rb\,\,=\,\,A\,;
\eeq
In the Wood-Saxon parametrization (see the last equation)
 $\rho_0 = 0.171\,(1/fm^3)$ \cite{WS}.

  In this approach, in order to calculate the hadron-nucleus cross sections, one  
needs to know the values of $\Delta_{\pom}$,$g$ and $G_{3\pom}$. 
 
 In this letter we will discuss the non-Glauber -type corrections to 
hadron-nucleus interaction. Our re-analysis of the problem is based on two
 recent achievements.

First, the Auger Collaboration has published the
 measurement of the proton-air total cross section at extremely high energies
 (W= 57\,TeV) with sufficiently small errors
 ($\sigma_{tot}\Lb \mbox{p - Air}\Rb\,\,=\,\,505 \pm 22
 \mbox{(stat)}\,{}^{+ 28}_{-36}\mbox{(syst)}\,mb$ \cite{AUGER}). 
 
 Second,  a model for high energy hadron-hadron scattering has been 
proposed which successfully 
 described LHC data, and which has included all  theoretical ingredients 
that have been found in QCD and N=4 SYM\cite{GLM}. The latter
 allows us to calculate the non-Glauber corrections, and to estimate the 
influence of these corrections on  the value of the
 proton-air cross section at very high energies.

In the next section we briefly review the main theoretical formulae for
 hadron-nucleus interactions. In section III we calculate the proton-air 
cross
 section using the 
model, all parameters of which have been fitted from the proton-proton
 data. In this section we estimate the difference between the 
Glauber-Gribov
 approach, and the alternative approach that includes the Pomeron 
interactions.
In the conclusions we summarize our results.

\section{Hadron-nucleus collisions}
\subsection{General approach}

\begin{figure}
\centerline{\epsfig{file=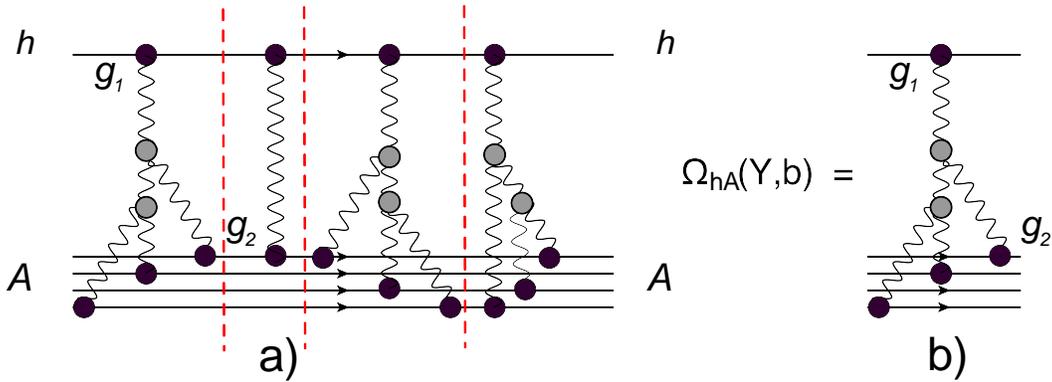,width=140mm}}
\caption{The  set of diagrams that contribute to the scattering amplitude of 
hadron-nucleus scattering in the kinematic region given by
 \eq{KRHA}. \fig{hAset}-b shows the hadron-nucleus irreducible
 diagrams while the general case is shown in \fig{hAset}-a. The
 vertical dashed lines indicate the hadron-nucleus states. The
 wavy lines denote the soft Pomerons. }
\label{hAset}
\end{figure}

In the kinematic region of \eq{KRHA}, the hadron-nucleus 
scattering amplitude can be written in an eikonal form in which the 
opacity $\Omega$ is given by sum of 
the 'fan' diagrams\cite{SCHW,BGLM} (see \fig{hAset}-b). 
\beq \label{HA1}
A_{\mbox{hA}}\Lb Y, b\Rb\,\,=
\,\,i \Lb 1\,\,-\,\,\exp \Lb - \frac{\Omega_{\mbox{hA}}\Lb Y; b\Rb}{2}\Rb \Rb, 
\eeq
with
\beq \label{HA2}
\Omega_{\mbox{hA}}\Lb Y; b\Rb \,\,=\,\,\frac{g_h\,g\,
G_{enh}(Y)\,S_A\Lb \vec{b}\Rb}{1 \,+\,g\,G_{3\pom}\,G_{enh}(Y)\,S_A\Lb \vec{b}\Rb}. 
\eeq
\par
For diagrams of \fig{hAset}, $G_{enh}(Y)$ is the Green's function of the 
Pomeron exchange which is equal to
\beq \label{GY}
G_{enh}(Y)\,\,\,=\,\,\,e^{\Delta_\pom\,Y}.
\eeq

From  \eq{HA1} and \eq{HA2}, we obtain that
\bea \label{HA3}
\sigma^{hA}_{tot}\,&=& 2 \int d^2 b\Lb 1\,\,-\,\,\exp \Lb - 
\frac{\Omega_{\mbox{hA}}\Lb Y; b\Rb}{2}\Rb \Rb; \nonumber\\
\sigma^{hA}_{el}\,&=&  \int d^2 b\Lb 1\,\,-\,\,\exp \Lb - 
\frac{\Omega_{\mbox{hA}}\Lb Y; b\Rb}{2}\Rb \Rb^2; \nonumber\\
\sigma^{hA}_{in}\,&=&  \int d^2 b\Lb 1\,\,-\,\,\exp \Lb - 
\Omega_{\mbox{hA}}\Lb Y; b\Rb\Rb \Rb.
\eea
The processes of diffractive production have been discussed in 
Refs.\cite{BGLM,BORY}.
\subsection{Main formulae}
\par
To describe the experimental data, we replace the Glauber-Gribov eikonal
formula by  \eq{HA1} and \eq{HA2}. 
  In addition,
we need to adjust this formula to our description of  
hadron-hadron data given in Ref.\cite{GLM}. In this model we introduce
 two additional ingredients that have not been included in   \eq{HA1}
 and \eq{HA2}.
\newline
1) A two channel Good-Walker model\cite{GW} which is 
exclusively responsible for low mass diffraction.
\newline
2) Enhanced Pomeron diagrams that lead to 
a different Pomeron Green's function. This mechanism provides 
the main contribution for high mass diffraction.

 The model we develop is based on  a two channel model which  takes into 
account  
the Good-Walker mechanism, in which   the observed physical 
hadronic and diffractive states are written in the form 
\beq \label{MF1}
\psi_h\,\,=\,\,\alpha\,\Psi_1+\beta\,\Psi_2\,;\,\,\,\,\,\,\,\,\,\,
\psi_D\,\,=\,\,-\beta\,\Psi_1+\alpha \,\Psi_2, 
\eeq
where $\alpha^2+\beta^2\,=\,1$. Note that Good-Walker diffraction 
is presented by a single wave function $\psi_D$.
 The vertex of the interaction of a Pomeron with a nucleon
 ($g$ in \eq{HA1} and \eq{HA2}) has in this model a more complex structure
$\tilde{g}\,\,=\,\,\alpha^2 g^{(1)} \,+\,\beta^2 g^{(2)}$. 
$g^{(k)}$  denotes the vertex of the Pomeron  
interaction with the state $k$ that have been described 
by either the wave functions $\Psi_1$ or $\Psi_2$.

\begin{figure}
\centerline{\epsfig{file=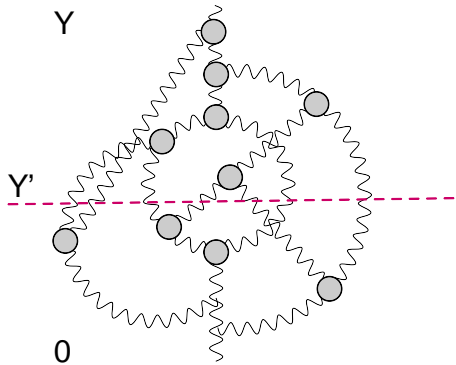,width=60mm}}
\caption{The  set of enhanced diagrams for the soft Pomeron. }
\label{enhdi}
\end{figure}

\begin{figure}
\centerline{\epsfig{file=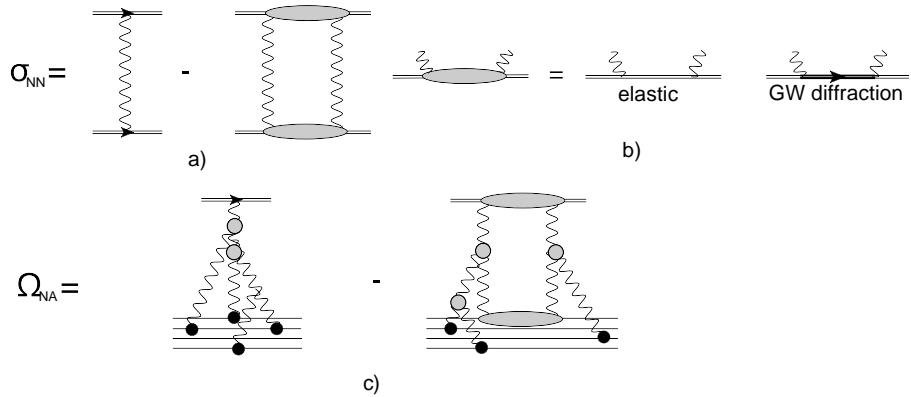,width=120mm}}
\caption{Proton-nucleus interaction (\fig{hAmod}-c)  in a simple model of \fig{hAmod}-a and \fig{hAmod}-b.}
\label{hAmod}
\end{figure}

 In our model \cite{GLM} we sum all enhanced diagrams for proton-proton
 scattering (see \fig{enhdi} and calculated $G_{enh}\Lb Y\Rb$ in the MPSI
 approximation \cite{MPSI}. We refer our reader to Ref.\cite{GLMAA} for the
 details of the calculations which lead to
\beq \label{GENY}
G_{enh} 
\left(Y\right)\,\,\,=\,\,\,\frac{1}{\gamma}\Big(1\,\,-\,\,\exp \Lb \frac{1}{T(Y)}\Rb\,
\frac{1}{T(Y)}\,\,\Gamma\Lb 0,\frac{1}{T(Y)} \Rb\Big)
\eeq
$\Gamma\Lb 0, x\Rb$ is the incomplete Gamma function 
(see {\bf 8.350 - 8.359} in Ref.\cite{RY})) ,
 $\gamma^2 \,\,\,=\,\,\int \,G^2_{3 \pom}\Lb k_{T,1}=0,k_T,k_T\Rb
 d^2 k_T,$ where $k_{T,i}$ are the transverse momenta of three Pomerons,
and $T\Lb Y\Rb$ is given by 
\beq \label{T}
T\Lb Y\Rb\,\,=\,\,\gamma\,e^{\Delta_\pom\,Y}
\eeq
The final formula that includes both the Good-Walker mechanism of low
 mass diffraction production, and the enhanced Pomeron diagrams, takes
 the following form:
\bea \label{FHA}
&&\sigma_{in}\Lb p+A; Y\Rb\,\,=\\
&&\int d^2 b \Lb 1 \,-\,\exp \Lb - 
\left\{ \sigma^{pp}_{tot} \frac{S_A(b)}{\Lb 1 \,+\,\tilde{g}\,G_{3 \pom}\,G_{enh}
\Lb Y\Rb \,S_A(b)\Rb} \,-\,(\sigma^{pp}_{el}\,+\,\sigma^{pp}_{diff}) 
\frac{S_A(b)}{\Lb 1 \,+\,\tilde{g}\,G_{3\pom}\,G_{enh}\Lb Y\Rb \,S_A(b\Rb)^2}\right\} \Rb 
\Rb. \nonumber
\eea
One can see that for $G_{3\pom}\,\to\,0$ \eq{FHA} reduces to Gribov-Glauber
 formula with one difference: instead of $\sigma_{in} \,= \,\sigma_{tot} -
 \sigma_{el} $ in the formula we have $\sigma_{in}\,=\,\sigma_{tot} - \sigma_{el}\,-\,\sigma_{dif}$ where $\sigma_{dif}\,=\,2 \sigma^{GW}_{sd} + \sigma_{dd}^{GW}$ as it was advocated in Ref.\cite{KOP}. Note, that the low mass of both single and double diffraction enters \eq{FHA}.
We consider  the simplest model for proton-proton interaction in  which 
 elastic and diffraction processes are taken into account (see \fig{hAmod}):
 the total cross section is given as a sum of one and two Pomeron exchanges(see \fig{hAmod}-a and \fig{hAmod}-b). 
  The Pomeron diagrams for this case, are shown in 
\fig{hAmod}-c.

Table 1 presents the parameters of our model\cite{GLM} that are needed
 to calculated the modified Gribov-Glauber formula of \eq{FHA}. For the
 scattering with air we use $S_{Air}\Lb b\Rb\,\,=\,\,0.78 S_{Ng}\Lb b \Rb
+ 0.22 S_{O}\Lb b \Rb$ where $Ng$ and $O$ stand for nitrogen and oxygen,
 respectively. For each of these nuclei we used the oscillatory
 parametrization, following  Ref. \cite{WS}.
 Our estimates will be discussed below.
\begin{table}
\begin{tabular}{|l|l|l|l|}
\hline \hline
$\Delta_\pom$ & $\tilde{g}( GeV^{-1})$ & $\gamma$ & $G_{3\pom}( GeV^{-1})$\\
\hline
 $0.23$ & 14.6 & 0.0045 & 0.03 \\
\hline \hline
\end{tabular}
\caption{Parameters of the model (see Ref.\cite{GLM} that are needed for 
the modified Gribov-Glauber formula of \protect\eq{FHA}}
\end{table}
\section{Comparison with the experiment}
     Before comparing with the experimental results, we would like to draw
the reader's attention to the fact that some of the experimental results 
shown might be overestimated, 
due to the possibility of the airshowers being created by helium nuclei, 
as well as protons.
 The importance of this phenomena has been investigated by Block\cite{BLO} 
and the Auger 
colloboration\cite{AUGER}. We refer the reader to these references for 
further details.
   All experiments, except Auger, assume a pure proton composition of the 
projectile. Auger concludes
that they have a contamination of about 25\% of helium, which produces an 
uncertainty of about
~ 30 mb (which is less than 10\% of their final result), and is included 
in their systematic error.

\begin{figure}
\centerline{\epsfig{file=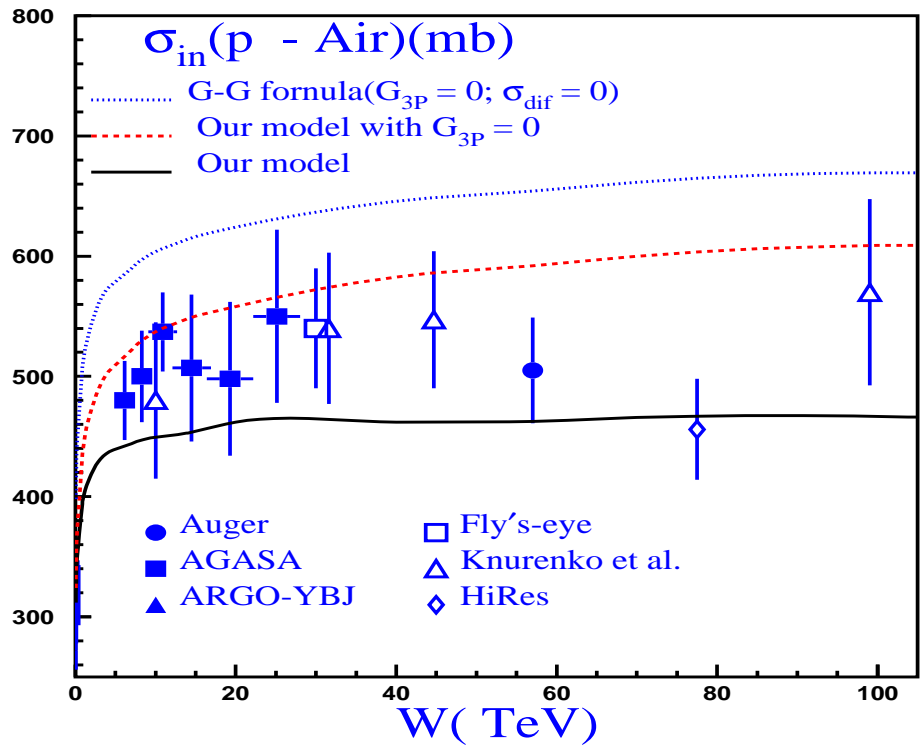,width=140mm,height=110mm}}
\caption{ Comparison the energy dependence of the total cross section for proton-Air interaction with the high energy experimental data. Data are taken from Refs.\cite{HRE,KNU,AGASA,EAS,ARGO}. }
\label{sig}
\end{figure}

The results of our calculations are shown in \fig{sig}. One can see that 
our model 
describes  the Auger and HiRes\cite{HRE} quite well. In \fig{sig} the two 
different 
simplifications of the exact formula of \eq{FHA} are plotted:
  Gribov-Glauber formula in which the interaction with the Pomerons are neglected
 ($G_{3\pom}=0$) as well as the cross section of  diffraction dissociation
 is considered to be small ($\sigma_{dif}=0$); and 
our   model with $G_{3\pom}=0$. The first lesson that we learn is that
 these three approaches produce  results whose difference are larger 
than
 the experimental errors of the Auger and HiRes  experiments.
 The second lesson is that our model gives a good description of the data
 at very high energies ($W = 57 \,TeV$ and $W=77.5 \,TeV$). Our total
 proton-proton cross section at $W = 57\,TeV$ is equal to 130 mb. It is
 instructive to note that all cosmic ray data lie between our two predictions:
 our modelwith $G_{3\pom}=0.03$, and our model with $G_{3\pom}=0$. Since the systematic errors are
 large
we cannot exclude the case with $G_{3\pom}=0$.
 However, this case has been eliminated by the
 LHC experiments where  high mass diffraction has been measured,

It should be stressed that our results show that the Pomeron interaction 
leads to a considerable contribution and cannot be neglected.
 We would like to stress that our model \cite{GLM} gives the
 smallest contribution for Pomeron interactions, in comparison
 with other attempts to describe the LHC data\cite{KMR,OST}.

\begin{figure}
\centerline{\epsfig{file=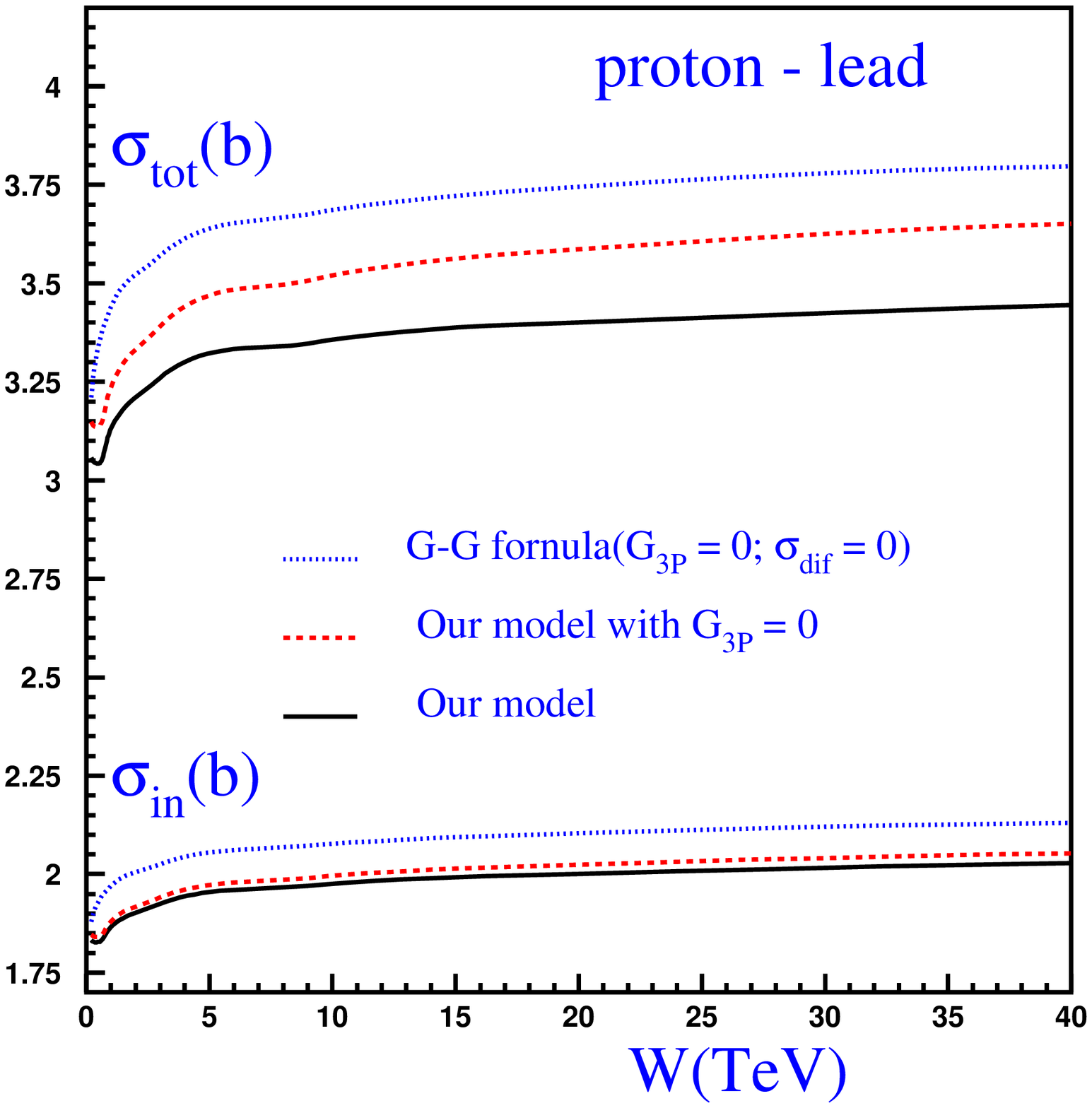,width=140mm,height=110mm}}
\caption{ Predictions of our model for proton-lead cross sections. In this figure $\sigma$ in barns while $W$ in TeV.  }
\label{sigpb}
\end{figure}

 \fig{sigpb} shows our predictions for the total and inelastic cross sections for proton-lead interaction at high energy.
 One can see that for heavy nuclei the difference between our approach and Gribov-Glauber formula is not large reaching about 11\% for total and 5\% for inelastic cross sections.It is instructive to note that the inelastic cross section for heavy nuclei is not sensitive to the Pomeron interactions and the major difference from Gribov-Glauber formula stems from Good-Walker mechanism for low mass diffraction in proton-proton collisions.
 
\section{Conclusions}
In this paper we show that  Pomeron interactions generate valuable
 corrections to Gribov-Glauber formula which is used for the extraction
 of proton-proton cross sections  from proton-air
 collisions at high energies. Bearing this in mind, we can attempt to use 
the 
cosmic ray data at
 ultra high energies ($W \,\,>\,\,1\,TeV$) to check our knowledge of very
 high energy phenomenology.

Although the cosmic ray experiments have large errors, we conclude
  that the Pomeron interaction and low mass diffraction
 production, have to be taken into account with all needed modifications
 of Gribov-Glauber formula for proton-air scattering.

Our model which describes all accelerator data at high energies also
 gives a good description of the cosmic ray data. Therefore, we
 conclude that our model \cite{GLM}
 can be used for ultra high energies.

  
  \section{Acknowledgements}
  
    This research was supported by the  Fondecyt (Chile) grant 1100648.

~

~

~

\end{document}